\documentstyle[aps,multicol,epsf]{revtex}

\begin{document}

\draft

\title{Supersymmetry and Localization in the Quantum Hall Effect} 

\author{J. Kondev\footnote{jane@barus.physics.brown.edu} and J.B. Marston}

\address{Department of Physics, Brown University, Providence, Rhode 
Island 02912-1843}

\date{February 28, 1997}

\bigskip
\maketitle
\centerline{cond-mat/9612223}

\widetext

\begin{abstract}
We study the localization transition in the integer quantum Hall 
effect as described by the network model of quantum percolation.  
Starting from a path integral representation of 
transport Green's functions for the network model, 
which employs both complex (bosonic) and Grassman (fermionic) 
fields, we map the problem of localization to the problem of  
diagonalizing a one-dimensional {\em non-Hermitian}  
Hamiltonian of interacting bosons and fermions.  An {\em exact} solution 
is obtained in a restricted subspace of the Hilbert space which 
preserves boson-fermion supersymmetry.
The physically relevant regime is investigated using the 
density matrix renormalization group (DMRG) method,
and critical behavior is found at the plateau transition. 
\end{abstract}


\newcommand{\bphi}{\mbox{\boldmath $\phi$}}
\newcommand{\bpsi}{\mbox{\boldmath $\psi$}}

\section{Introduction}
\label{sec:intro}
The physics of electronic transport in the presence of impurities
is rich and incompletely understood.  Disorder plays a crucial role
in the integer quantum Hall effect (IQHE) as it leads to plateaus in the 
Hall resistance as a function of applied magnetic field.  Transitions 
between plateaus are an especially interesting critical phenomenon.  
Critical behavior was predicted by Levin, Libby, and Pruisken\cite{Pruisken} 
and was observed experimentally
by Wei et al.\cite{Tsui} for temperatures near the critical point 
at absolute zero, and by Koch et al.\cite{Koch} for different
sample widths.  The correlation length exponent $\nu$ was 
measured to be approximately $\nu \approx 2.3$ and the dynamical exponent
$z$ was found to be equal to unity. 

Progress toward a theoretical understanding of the plateau transition was made
with the introduction of a quantum tunneling network model by Chalker and
Coddington\cite{Chalker}.  Subsequent numerical studies\cite{numerics} of the 
Chalker-Coddington network model yielded values for $\nu$ consistent with
experiment.  An argument based on the semi-classical picture of electron
motion in the IQHE suggests that the value of $\nu$ is $7/3$ 
\cite{rus}, but so far analytical derivations are still unconvincing.  
Lee\cite{Dung-Hai} mapped the network model to the $n \rightarrow 0$ replica
limit of a one-dimensional quantum $SU(2n)$ Heisenberg antiferromagnet, where
the plateau transition is equivalent to a transition between two dimerized
phases. Lee and Wang\cite{Ziqiang} extended this mapping to the replica limit 
of an associated Hubbard model. 

In a continuum picture Ludwig {\em et al.}\cite{Andreas} 
mapped the IQHE problem  to a model of two-dimensional Dirac fermions 
interacting via random scalar and gauge fields.  
They speculated that the fixed point controlling the plateau transitions was 
at strong-coupling and thus not directly accessible by  methods of 
weak-coupling perturbation theory or bosonization. 

There is another field theory method besides replication which permits the
averaging over quenched disorder: supersymmetry.  McKane showed that the
introduction of equal numbers of bosonic and fermionic degrees of freedom made
the partition function independent of the particular realization of disorder.
Disorder-averaged observables then could be expressed as the Green's functions
of a field theory\cite{McKane}.  Efetov used this idea to reformulate the 
$\sigma$-model approach to Anderson localization\cite{Efetov} and 
Affleck, in a saddle-point calculation\cite{Ian}, showed that the density of
states in a uniform magnetic field and a white noise potential was 
non-vanishing.  Zirnbauer\cite{Zirnbauer} applied the method to transport
properties and derived both the supersymmetric
$\sigma$-model and an antiferromagnetic supersymmetric spin-chain. 
Balents, Fisher and Zirnbauer\cite{Balents} recently
applied this approach to the simpler problem of the chiral metal and 
exactly solved the resulting ferromagnetic supersymmetric quantum spin-chain. 
Magnon excitations were identified as the expected diffusive modes.

In contrast to the supersymmetric ferromagnet\cite{Balents,subir}, 
little is known about the antiferromagnetic supersymmetric spin chain.  In
this paper we rederive it in terms familiar to condensed matter physicists
and find an exact analytic solution in a particular limit of the model, which
opens up the possibility of studying the plateau transition perturbatively.   
We also examine the chain in the physically relevant limit by means of the 
Density Matrix Renormalization Group (DMRG) method and demonstrate the 
existence of a critical point. The correlation length exponent that we
find is consistent with previous numerical studies of the Chalker-Coddington
network model.

\section{Network model}
\label{sec:network}
\begin{figure}
\center
\epsfxsize = 10cm \epsffile{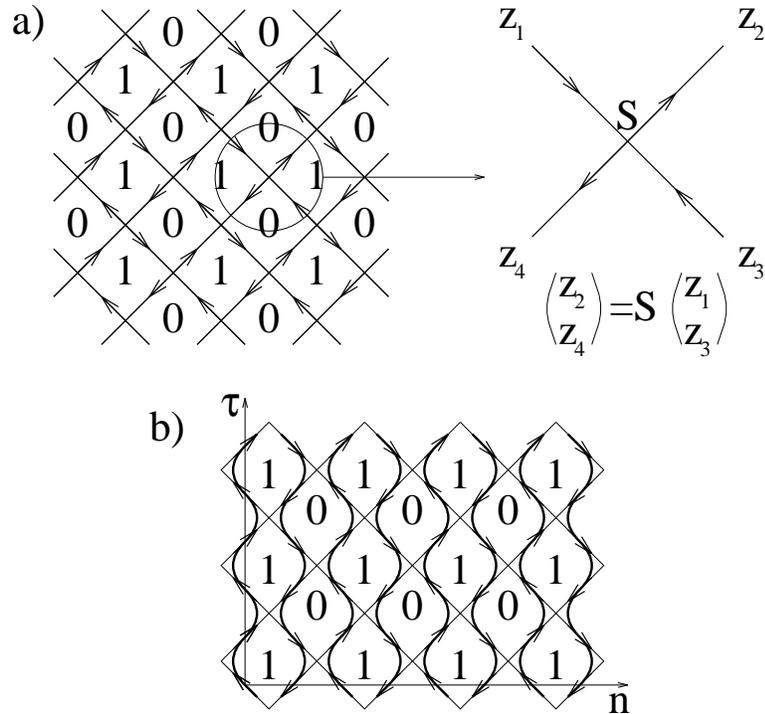}
\caption{
\label{network_fig}
(a) The Chalker-Coddington network model. Amplitudes of the electron 
wavefunction (${\rm z}_i$) are assigned to bonds, while the {\em random}
scattering matrices (S) are assigned to vertices of the network.  1 and 
0 denote Hall droplets associated with  two consecutive plateaus.  
(b) The Hamiltonian representation of the network model with 
counterpropagating chiral fermions as the edge states of the Hall droplets; 
the tunneling between the edge states is random. 
}
\end{figure}

To describe the transition between quantum Hall plateaus Chalker and Coddington
introduced a lattice model that combines the two most important features 
of electronic transport in a magnetic field and in the presence of 
disorder: percolation and quantum tunneling. The model consists of  a  
network of tunneling centers which form a square lattice in the x-y plane;
Fig.~\ref{network_fig}(a). The electron 
wavefunction is represented by complex amplitudes ${\rm z}_i$ which are
defined at the bond centers.
In the semiclassical picture of electron transport in the IQHE \cite{Trugman} 
the guiding center of the electron's orbit propagates along a  
contour line of the random potential.  At the saddle points, where two 
contours pass within a cyclotron radius, the electron can tunnel.
Due to the presence of a perpendicular magnetic field the electron acquires a 
random Aharonov-Bohm phase between two tunneling events.  
In the network model this phase is represented by random 
scattering matrices (S) that relate the amplitudes of the electron 
wavefunction on the outgoing bonds to the amplitudes on the incoming 
bonds, Fig.~\ref{network_fig}(a).  The scattering 
matrices are unitary and are typically chosen to be independently  
Gaussian distributed. 

An alternative description of the network model was recently proposed by 
Lee\cite{Dung-Hai} where the random scattering matrices
are replaced by a Hamiltonian which describes a one-dimensional array 
of counterpropagating edge states in the presence of disorder, as shown in
Fig.~\ref{network_fig}(b). The direction of propagation will be later 
interpreted as the imaginary time; to emphasize this point of view
we relabel the $y$-coordinate as $\tau$ and also denote the 
$x$-coordinate with the integer index $n$.        

Expressed in terms of electron creation 
$\hat{\psi}^\dagger$ and annihilation $\hat{\psi}$
operators the Hamiltonian may be written as 
\begin{equation}
\label{hamilton}
H[\hat{\psi}^\dagger, \hat{\psi}]
= \sum_{n,\tau} \ [(-1)^n \hat{\psi}_{n,\tau}^\dagger 
({\rm i} \partial_\tau + w_{n, \tau}) \hat{\psi}_{n,\tau} 
- (t_{n,\tau} \hat{\psi}_{n,\tau}^\dagger \hat{\psi}_{n+1,\tau}
+ t_{n,\tau}^* \hat{\psi}_{n+1,\tau}^\dagger \hat{\psi}_{n,\tau})], 
\end{equation}
where the sum is over the discrete vertices of the network model, and 
$\hat{\psi}_{n,\tau}^\dagger({\rm i} \partial_\tau) 
\hat{\psi}_{n,\tau}$ is defined in momentum (Fourier) space, in the limit of
small $k_\tau$, as 
$\hat{\psi}_{n,k_\tau}^\dagger (k_\tau) \hat{\psi}_{n,k_\tau}$. 
To avoid ordering ambiguities, 
for now we keep $\tau$ discrete and refrain from taking the continuum limit 
until later, after the average over disorder has been performed.
The Hamiltonian, Eq.~\ref{hamilton},
reflects the chiral nature and linear dispersion of the edge states, 
with alternating propagation
forward and backward in the $\tau$ direction at Fermi velocities $\pm 1$. 
It also incorporates complex tunneling amplitudes $t_{n,\tau}$  
with random phases, which represent the random Aharonov-Bohm phases of the 
electrons discussed above, as well as random on-site potentials $w_{n,\tau}$.
The random terms are
taken to be independently Gaussian distributed with zero mean, and variance
\begin{eqnarray}
\label{disorder}
\ll w_{n, \tau}~ w_{n^\prime, \tau^\prime} \gg 
&=& 2 U~ \delta_{n, n^\prime}~ \delta_{\tau, \tau^\prime} 
\nonumber \\ 
\ll t_{n, \tau}^*~ t_{n^\prime, \tau^\prime} \gg &=& 
J_{n}~ \delta_{n, n^\prime}~ \delta_{\tau, \tau^\prime} \ , 
\ \ \ J_n = J~ \times~ (1+(-1)^n R)\ .
\end{eqnarray}
Staggered modulation in the parameter $J_n$ accounts for differences in 
tunneling between counterpropagating electrons in 
adjacent plaquette-columns as depicted in Fig.~\ref{network_fig}(b). 
As $R\rightarrow -1$  electrons only circulate clockwise around the $1$
plaquettes.  We can picture these plaquettes as isolated lakes of  
electrons (regions of attractive potential with filling fraction $1$) 
separated by dry land (repulsive potential and filling fraction $0$).  In
the opposite limit $R \rightarrow 1$ electrons circulate counterclockwise
only around $0$ plaquettes, which corresponds 
to isolated islands (empty regions of filling fraction $0$) in the midst of
an ocean of (filling fraction $1$) electrons.  Precisely at  
$R = 0$ the land and sea 
cover equal areas and we expect critical quantum percolation associated
with the transition between the 0 and 1 plateau.  
The percolating
edge states are characterized by a diverging length scale, 
$\xi \sim R^{-\nu}$.
Critical behavior in accord with this scenario was found from the  
disorder averaged   
transport Green's function of  the Hamiltonian (Eq.~\ref{hamilton}) by 
Wang and Lee\cite{Ziqiang}. From a Monte-Carlo treatment of the replicated 
Hamiltonian they obtained a correlation length exponent 
$\nu = 2.33 \pm 0.03$, in good agreement with experiments and 
other numerical simulations of the network model.

\section{Transport Green's functions}
\label{sec:transport}

To study the localization-delocalization transition in the network model as  
described by the random Hamiltonian, Eq.~\ref{hamilton},
we focus on the disorder-averaged transport Green's function\cite{stone} 
\begin{equation}
\label{tGfun}
K(1,2) = \ll G_+(1,2)~ G_-(2,1) \gg   
\end{equation}
where 
\begin{equation}
\label{raGfun}
G_{\pm}(1,2) = \langle 1| \frac{1}{H-E\pm {\rm i}\eta} |2\rangle 
\end{equation}
are the advanced and retarded Green's functions; $E$ is the energy and 
$\eta$ a positive infinitesimal.  We introduce the usual
shorthand for coordinates with $1 \equiv (n_1, \tau_1)$ and 
$2 \equiv (n_2, \tau_2)$.  In the localized regime, $R$ is non-zero, and 
we expect the transport Green's function to be short-ranged
\begin{equation}
\label{shortrange}
K(1,2) \sim \exp(-r_{12}/\xi)
\end{equation}
where $r_{12}$ is the distance between points 1 and 2, and $\xi \sim R^{-\nu}$.
The correlation length $\xi$ 
reflects the finite extent of the electron wavefunction. 
To calculate the matrix elements of the resolvent of $H$ appearing in 
the expression for the advanced and retarded Green's functions, 
Eq.~\ref{raGfun}, we introduce a pair of complex scalar fields 
\begin{equation}
\label{compair}
\bphi(n,\tau) = \left( \begin{array}{c} \phi_\uparrow(n,\tau) \\
                                        \phi_\downarrow(n,\tau) 
                       \end{array} \right) \ .
\end{equation}
The Green's functions\cite{stone} can  now be written as  
\begin{eqnarray}
\label{pathint}
G_+(1,2) &=& -{\rm i}~ \langle \phi_\uparrow(1) \phi^*_\uparrow(2) \rangle 
\nonumber \\
G_-(1,2) &=& +{\rm i}~ \langle \phi_\downarrow(1) \phi^*_\downarrow(2) 
\rangle \ ,    
\end{eqnarray} 
where averages of observables ${\cal O}$
are taken with respect to a Gaussian measure,
\begin{eqnarray}
\label{gauss}
\langle {\cal O} \rangle &=& {\cal Z}^{-1} \int {\rm D}\bphi {\rm D}\bphi^*~ 
{\cal O}[\bphi, \bphi^*]~ \exp(-S[\bphi,\bphi^*])  
\nonumber \\
S[\bphi, \bphi^*] &=& - {\rm i} \bigg{\{} H[\phi_\uparrow] + 
{\rm i} \eta \sum_{n,\tau} \phi_\uparrow(n,\tau) \phi^*_\uparrow(n,\tau) 
\bigg{\}}    
+ {\rm i} \bigg{\{} H[\phi_\downarrow] - {\rm i} \eta \sum_{n,\tau} 
\phi_\downarrow(n,\tau) \phi^*_\downarrow(n,\tau) \bigg{\}}\ .
\end{eqnarray}
In the above equation we have set $E = 0$; in the advanced and retarded 
Green's functions non-zero energy is easily recovered by replacing $\eta$ with 
$\eta \pm {\rm i}E$. The choice of signs in the terms 
proportional to $\eta$ is 
dictated by the requirement that the Gaussian integrals be convergent
for $\eta > 0$.  The quadratic form $H[\phi]$ is 
obtained simply by replacing the {\em operators} $\hat{\psi}_{n,\tau}$ and
$\hat{\psi}_{n,\tau}^\dagger$ in the Hamiltonian of the network model, 
Eq.~\ref{hamilton}, with complex {\em scalar fields} $\phi(n,\tau)$ and 
$\phi^*(n,\tau)$.  Finally, the normalization factor
\begin{equation}
\label{norm}
{\cal Z} = \int {\rm D}\bphi {\rm D}\bphi^* \exp(-S[\bphi,\bphi^*])
\end{equation} 
can be written as Gaussian integral over {\em Grassman fields}\cite{negele} 
\begin{equation}
\label{normGrass}
{\cal Z}^{-1} = \int {\rm D}\bpsi {\rm D}\bar{\bpsi} 
\exp(-S[\bpsi, \bar{\bpsi}]) 
\end{equation}
where the anticommuting fields
\begin{equation}
\label{Graspair}
\bpsi(n,\tau) = \left( \begin{array}{c} \psi_\uparrow(n,\tau) \\
                                        \psi_\downarrow(n,\tau) 
                       \end{array} \right)
\end{equation}
are the supersymmetric partners of the scalar fields.  Using both
Eqs.~\ref{normGrass} and~\ref{gauss}, 
the correlation functions, Eq.~\ref{pathint}, may be 
written as a combined Gaussian integral over scalar and Grassman fields
\begin{equation}
\label{Gaussfinal}
\langle {\cal O} \rangle = \int {\rm D}\bphi {\rm D}\bphi^* {\rm D}\bpsi 
{\rm D}\bar{\bpsi}~ {\cal O}[\bphi, \bphi^*]~ 
\exp(- S[\bphi,\bphi^*] - S[\bpsi,\bar{\bpsi}]) .
\end{equation}
The utility of this representation lies in the fact that, for each realization
of disorder, the normalization factor has been absorbed into the action. 
Now the average over disorder can be interchanged with integration over
the scalar and Grassman fields, and the problem of computing the
disorder-averaged transport Green's function, Eq.~\ref{tGfun}, 
reduces to the problem of calculating correlation functions in a {\em pure} 
field theory, albeit one with interacting fermionic and bosonic degrees 
of freedom.
Our strategy in what follows will be to perform the disorder average in the 
path integral, Eq.~\ref{Gaussfinal}. Then, because the $\tau$-derivative term
is linear, the disorder-averaged path integral will be interpreted 
as a coherent state path integral for a {\em one-dimensional} 
Hamiltonian. The problem of calculating disorder-averaged correlation 
functions for the network model which is {\em infinite} in the 
$\tau$ direction 
is thus transformed to the problem of computing expectation values 
in the {\em ground state}  of this Hamiltonian.   

\section{Non-Hermitian Hamiltonian}
\label{sec:non-herm}
The supersymmetry method, as discussed in the preceding section, allows us
to perform the average over disorder {\em inside} the path integral. 
Before proceeding with this step we first transform the 
scalar and Grassman fields on even sites, 
\begin{equation}
\label{transfeven}
\mbox{n even:} \ \ \ \  
\begin{array}{l}\phi_\downarrow(n,\tau) \to \phi^*_\downarrow(n,\tau)\\ \\
                       \phi^*_\downarrow(n,\tau) \to \phi_\downarrow(n,\tau) 
        \end{array}  \ \ \ 
\begin{array}{l}\psi_\downarrow(n,\tau) \to 
                                      -\bar{\psi}_\downarrow(n,\tau)\\ \\
                       \bar{\psi}_\downarrow(n,\tau) \to 
                                      \psi_\downarrow(n,\tau) 
        \end{array} 
\end{equation}
and on odd sites 
\begin{equation}
\label{transfodd}
\mbox{n odd:} \ \ \ \  
\begin{array}{l}\phi_\uparrow(n,\tau) \to \phi^*_\uparrow(n,\tau)\\ \\
                       \phi^*_\uparrow(n,\tau) \to \phi_\uparrow(n,\tau) 
        \end{array}  \ \ \ 
\begin{array}{l}\psi_\uparrow(n,\tau) \to 
                                      -\bar{\psi}_\uparrow(n,\tau)\\ \\
                       \bar{\psi}_\uparrow(n,\tau) \to 
                                      \psi_\uparrow(n,\tau) \; ; 
        \end{array}   
\end{equation}
the remaining fields are left unchanged.
The transformation is guided by two requirements.  First, the linear 
derivative term in the action acquires the canonical form for both  
scalar and Grassman fields \cite{negele},
\begin{equation}
\label{derivative}
S_\tau = \sum_{n,\tau} \bigg{[}
       \bar{\bpsi}(n,\tau) \cdot \partial_\tau \bpsi(n,\tau) + 
       \bphi^*(n,\tau) \cdot \partial_\tau \bphi(n,\tau) \bigg{]}\ .
\end{equation}   
Second, the sign of the $\eta$ term in the bosonic
part of the action must be such to ensure convergence of the path integral; see
Eq.~\ref{Sterms} below. 

Next we perform the disorder average inside the path integral 
using Eq.~\ref{disorder}:
\begin{equation}
\label{disave}
\ll \exp(-S[\bphi,\bphi^*]-S[\bpsi,\bar{\bpsi}]) \gg  
~ =~ \exp(-S_{\rm SUSY}[\bphi,\bphi^*,\bpsi,\bar{\bpsi}]) \ . 
\end{equation}
The action $S_{\rm SUSY}$ can be written as a sum of three terms
\begin{equation}
\label{Ssusy}
S_{\rm SUSY}  =  S_\tau + S_U + S_J + S_\eta 
\end{equation}
where $S_\tau$ is given by Eq.~\ref{derivative}, and
\begin{eqnarray} 
\label{Sterms}
S_U &=& U~ \sum_{n, \tau} [\bar{\bpsi}(n, \tau) \sigma_z \bpsi(n, \tau) 
        + \bphi^*(n, \tau) \sigma_z \bphi(n, \tau)]^2 \nonumber \\
S_J &=& \sum_{n, \tau} J_n~ [(-1)^{n+1} {\cal C}(n, \tau) + {\cal B}(n,\tau)] 
          [(-1)^n \bar{{\cal C}}(n, \tau) + {\cal B}^*(n, \tau)] \nonumber \\
S_\eta &=& \eta~ \sum_{n, \tau} [\bar{\bpsi}(n, \tau) \cdot \bpsi(n, \tau) + 
             \bphi^*(n, \tau) \cdot \bphi(n, \tau)] \ .  
\end{eqnarray} 
$\sigma_z = \mbox{diag}(1,-1)$ is the $2 \times 2$ Pauli matrix operating in 
spin space ($\uparrow$, $\downarrow$), and
we have introduced new fields bilinear in the 
original Grassman and scalar fields which live on the bonds,
\begin{eqnarray}
\label{bilinear}
{\cal B} (n, \tau) &=& 
               \phi^*_\uparrow(n,\tau)\phi^*_\uparrow(n+1,\tau) - 
               \phi_\downarrow(n,\tau)\phi_\downarrow(n+1,\tau) 
\nonumber \\
{\cal C} (n, \tau) &=& 
             \bar{\psi}_\uparrow(n,\tau) \bar{\psi}_\uparrow(n+1,\tau)
             + \psi_\downarrow(n,\tau) \psi_\downarrow(n+1,\tau)\ .  
\end{eqnarray} 
Now it is safe to take the $\tau$-continuum limit and interpret the 
$\tau$ direction as imaginary time. 
The action $S_{\rm SUSY}$ is precisely that which appears in 
a coherent state path integral representation of the 
partition function for a one-dimensional Hamiltonian given by:
\begin{equation}
\label{Hsusy}
H_{\rm SUSY} =  H_U + H_J + H_\eta \ . 
\end{equation}
The different terms in the Hamiltonian can be read off from the corresponding
terms in $S_{\rm SUSY}$ {\em only after} the scalar and Grassman fields have
been properly ordered: the fields are commuted and anticommuted so that 
$\phi^*$ and $\bar{\psi}$ always appear before $\phi$ and 
$\psi$. The normal ordering is essential\cite{negele} 
because the scalar and Grassman fields appear
in the coherent state path integral as {\em left and right} eigenvalues of 
{\em creation and annihilation} operators, respectively.  
Introducing bosonic operators which satisfy the usual canonical commutation 
relations  
$[\hat{b}_\uparrow,~ \hat{b}^\dagger_\uparrow] 
= [\hat{b}_\downarrow,~ \hat{b}^\dagger_\downarrow] = 1$ 
and fermionic operators which satisfy canonical anticommutation relations  
$\{ \hat{c}_\uparrow,~ \hat{c}^\dagger_\uparrow \} 
= \{ \hat{c}_\downarrow,~ \hat{c}^\dagger_\downarrow \}
= 1$, normal ordering, and dropping the hat over the operators for clarity, 
we obtain
\begin{eqnarray}
\label{Hterms}
H_U &=& U~ \sum_n [({\cal N}^c_{\uparrow n} - {\cal N}^c_{\downarrow n} 
+ {\cal N}^b_{\uparrow n} - {\cal N}^b_{\downarrow n})^2 
- ({\cal N}^c_{\uparrow n} + {\cal N}^c_{\downarrow n}  
+ {\cal N}^b_{\uparrow n} + {\cal N}^b_{\downarrow n})] 
\nonumber \\
H_J &=& \sum_n J_n \: [ : {\cal B}_n {\cal B}^\dagger_n : - : {\cal C}_n 
{\cal C}^\dagger_n : + (-1)^{n+1}({\cal C}_n {\cal B}^\dagger_n -       
{\cal C}^\dagger_n {\cal B}_n)] 
\nonumber \\
H_\eta &=& \eta~ \sum_n [{\cal N}^b_{\uparrow n} + {\cal N}^b_{\downarrow n} 
+ {\cal N}^c_{\uparrow n} + {\cal N}^c_{\downarrow n}]\ .
\end{eqnarray}
We have introduced the boson and fermion number operators
${\cal N}^b_\uparrow = b^\dagger_\uparrow b_\uparrow$, 
${\cal N}^b_\downarrow = b^\dagger_\downarrow b_\downarrow$, 
${\cal N}^c_\uparrow = c^\dagger_\uparrow c_\uparrow$, 
and ${\cal N}^c_\downarrow = c^\dagger_\downarrow c_\downarrow$. 
The remaining bilinear operators take their form from Eq.~\ref{bilinear}:
\begin{eqnarray}
\label{bilinearII}
{\cal B}_n &=& b^\dagger_{\uparrow n} b^\dagger_{\uparrow n+1} -    
              b_{\downarrow n} b_{\downarrow n+1} 
\nonumber \\
{\cal C}_n &=& c^\dagger_{\uparrow n} c^\dagger_{\uparrow n+1} +    
              c_{\downarrow n} c_{\downarrow n+1} \ , 
\end{eqnarray}
and products of these operators which appear in $H_J$ are normal ordered. 
The terms $H_U$ and $H_\eta$ are Hermitian while $H_J$ is {\em not}.
The lack of Hermiticity has its origin in our treatment of one of the spatial
directions as the time direction.  Evolution forward in real time must be
unitary, but no such restriction holds for evolution in a spatial direction. 

\section{Exactly solvable limit}
\label{sec:infiniteU}

The ground state of $H_{\rm SUSY}$ holds the key to understanding the 
localization transition in the  network model. The disorder-averaged 
transport Green's function can be written as a ground state average, 
\begin{equation}
\label{tr=gr}
K(1,2) = \langle c^\dagger_\uparrow(1) c^\dagger_\uparrow(2) 
c^\dagger_\downarrow(2) c^\dagger_\downarrow(1)\rangle_{\rm SUSY}  
\end{equation} 
if $1$ is an odd lattice site and $2$ is an even site.  This equation 
follows from Eqs.~\ref{tGfun} and~\ref{pathint}, once the transformations
in Eqs.~\ref{transfeven} and~\ref{transfodd} have been taken into account,
and after the Grassman fields have been replaced with the fermion creation 
and annihilation operators in the Heisenberg representation.
To find the ground state we first perform a canonical transformation   
\begin{equation}
\label{cantra}
   \begin{array}{l} c_\downarrow \to c^\dagger_\downarrow \\ \\
                    c^\dagger_\downarrow \to c_\downarrow 
        \end{array} \ \ \ \ \ 
   \begin{array}{l} b_\downarrow \to \bar{b}_\downarrow \equiv 
	b^\dagger_\downarrow \\ \\
                    b^\dagger_\downarrow \to -b_\downarrow 
        \end{array}
\end{equation}
on every site of the chain.  This transformation preserves the canonical 
commutation and anticommutation relations
between creation and annihilation operators, and it 
ensures that $H_{\rm SUSY}$ commutes with the number operator 
${\cal N}_n={\cal N}^c_{\uparrow n}+{\cal N}^c_{\downarrow n}+
{\cal N}^b_{\uparrow n}+{\cal N}^b_{\downarrow n}$ on each site, 
where the new ${\cal N}^b_{\downarrow} = \bar{b}_\downarrow b_\downarrow$. 
Note, however, that in the bosonic sector the new $b_\downarrow$ and 
$\bar{b}_\downarrow$ are {\em not} Hermitian conjugates of each other,  
rather, $(\bar{b}_\downarrow)^\dagger = - b_\downarrow$. 
The transformation also removes
the explicit $SU(2)$ spin symmetry as the $\uparrow$-spin
operators are unchanged.  

As $[H_{\rm SUSY}, {\cal N}_n] = 0$ the Hilbert space breaks up into sectors
with different numbers of particles on each site, and we are faced with the 
task of identifying the lowest energy sector. After performing the canonical 
transformation, Eq.~\ref{cantra}, the new $H_U$ term is 
\begin{equation}
\label{newU}
H_U =  U~ \sum_n [({\cal N}^b_{\uparrow n} + {\cal N}^b_{\downarrow n} 
+ {\cal N}^c_{\uparrow n} + {\cal N}^c_{\downarrow n})^2 
- ({\cal N}^b_{\uparrow n} - {\cal N}^b_{\downarrow n}  
+ {\cal N}^c_{\uparrow n} - {\cal N}^c_{\downarrow n})]\ . 
\end{equation}
For large $U$ the ground state 
lies in the subspace of the total Hilbert space which 
has either no particles, or just one $\uparrow$-fermion or 
one $\uparrow$-boson per site, 
as there is a relative energy cost of $2 U$ for a $\downarrow$-particle. 
The energy of the vacuum state, however, is raised with respect to the 
singly-occupied states once the other two terms in the Hamiltonian, 
Eq.~\ref{Hterms} are included.  As the ground state occupancy is thus 
non-zero, the density of states (DOS) is also non-zero.  This is as expected
because the DOS does not vanish either at the Anderson transition\cite{stone}
or in the Landau level problem\cite{Ian}.

For large $U$ the supersymmetric exchange Hamiltonian ($H_J$) 
can be written as
\begin{equation}
\label{newJ}
H_J  = \sum_n J_n~ \big{[} {\cal N}^b_{\uparrow n} {\cal N}^b_{\uparrow {n+1}} 
- {\cal N}^c_{\uparrow n} {\cal N}^c_{\uparrow n+1} + (-1)^{n+1}  
(c^\dagger_{\uparrow n} b_{\uparrow n} 
c^\dagger_{\uparrow n+1} b_{\uparrow n+1} + 
b^\dagger_{\uparrow n} c_{\uparrow n} 
b^\dagger_{\uparrow n+1} c_{\uparrow n+1}) \big{]}\ ;
\end{equation}   
all the terms with $\downarrow$-spin fermions or bosons have been eliminated. 
Physically this simply means that there is no transport in the $U\gg J$
limit.

The large-$U$ Hamiltonian, Eq.~\ref{newJ}, is non-Hermitian, degenerate,
and it is also 
{\em defective}: its eigenstates do not span the Hilbert space.   
The defect in $H_J$ is evident even for just two sites with a four-dimensional
Hilbert space spanned by:
$\{ b^\dagger_{\uparrow 1} b^\dagger_{\uparrow 2} | 0 \rangle,~ 
b^\dagger_{\uparrow 1} c^\dagger_{\uparrow 2} | 0 \rangle,~ 
c^\dagger_{\uparrow 1} b^\dagger_{\uparrow 2} | 0 \rangle,~ 
c^\dagger_{\uparrow 1} c^\dagger_{\uparrow 2} | 0 \rangle \}$. 
Only three (not four) distinct eigenstates of $H_J$ can be constructed.
Dropping the spin index for clarity, and setting $J_n=1$,
the two site Hamiltonian may be written:
\begin{equation}
H_J = b^\dagger_2 b_2 b^\dagger_1 b_1 - c^\dagger_2 c_2 c^\dagger_1 c_1
- c^\dagger_2 c^\dagger_1 b_1 b_2 + b^\dagger_2 b^\dagger_1 c_1 c_2~ . 
\label{twosite}
\end{equation}
As the Hamiltonian is non-symmetric, there are both right eigenstates:
\begin{equation}
H_J~ | \Psi^R(a) \rangle = \epsilon_a~ | \Psi^R(a) \rangle	
\label{righteigenstates}
\end{equation}   
and left eigenstates:
\begin{equation}
\langle \Psi^L(a) |~ H_J = \langle \Psi^L(a) |~ \epsilon_a~ , 
\label{lefteigenstates}
\end{equation}   
with $\langle \Psi^L(a) | \equiv (| \Psi^L(a) \rangle)^\dagger \neq \langle
\Psi^R(a) |$ in general.
In this case the three eigenstates are degenerate, with zero energy.  The right
eigenstates are: 
\begin{eqnarray}
| \Psi^R(1) \rangle &=& c^\dagger_1 b^\dagger_2 | 0 \rangle
\nonumber \\
| \Psi^R(2) \rangle &=& b^\dagger_1 c^\dagger_2 | 0 \rangle
\nonumber \\
| \Psi^R(3) \rangle &=& (b^\dagger_1 b^\dagger_2 
+ c^\dagger_1 c^\dagger_2) | 0 \rangle
\label{twositestates}
\end{eqnarray}
as may be easily verified by direct substitution into 
Eq.~\ref{righteigenstates}.
The defect may be removed with the addition of a small, supersymmetry breaking,
chemical potential for the fermions:   
\begin{equation}
H_J \rightarrow H_J + \mu_c N_c;\ \mu_c < 0
\label{chempot}
\end{equation}
where $N_c \equiv \sum_n ({\cal N}^c_{\uparrow n} 
+ {\cal N}^c_{\downarrow n})$ is the total fermion number operator.
This term lifts some of the degeneracy and splits the third state into 
two distinct ones:
\begin{equation}
| \Psi^R(3 \pm) \rangle = {{1}\over{\sqrt{N}}}~
\big{[} (-\mu_c \pm \sqrt{\mu_c^2 - 2 \mu_c}) b^\dagger_1 b^\dagger_2 
+ (\mu_c \pm \sqrt{\mu_c^2 - 2 \mu_c}) 
c^\dagger_1 c^\dagger_2 \big{]} | 0 \rangle
\label{defectivestates}
\end{equation}
with energies $\epsilon_{3 \pm} = \mu_c \pm \sqrt{\mu_c^2 - 2 \mu_c}$; $N$
is the normalization constant.  The
leading dependence of the energies on $\sqrt{\mu_c}$, as opposed to 
linear dependence on $\mu_c$, is a telltale sign of  
the defectiveness of the Hamiltonian in the $\mu_c \rightarrow 0$
limit.  The left eigenstates are easily found:
\begin{eqnarray}
\langle \Psi^L(1) | &=& \langle 0 | b_2 c_1   
\nonumber \\
\langle \Psi^L(2) | &=& \langle 0 | c_2 b_1   
\nonumber \\
\langle \Psi^L(3 \pm) | &=& \langle 0 | \big{[} 
(-\mu_c \pm \sqrt{\mu_c^2 - 2 \mu_c}) b_2 b_1 
- (\mu_c \pm \sqrt{\mu_c^2 - 2 \mu_c}) c_2 c_1 \big{]}~ {{\pm 1}\over{\sqrt{N}}}
\label{leftstates}
\end{eqnarray}
Now the four states can be properly normalized, 
\begin{equation}
\langle \Psi^L(b) | \Psi^R(a) \rangle = \delta_{ab}, 
\label{normalization}
\end{equation}
by setting $N = -4 \mu_c \sqrt{\mu_c^2 - 2 \mu_c}$.  

Before turning to the problem of more than two sites, we examine the 
eigenstates of $H_{\rm SUSY}$ with $\downarrow$-spin excitations 
in the physical, finite-$U$ limit. 
By direct examination of the Hamiltonian we find that eight 
right-eigenstates can be formed by placing one $\uparrow$-spin and one 
$\downarrow$-spin particle on the two sites: 
\begin{equation}
\label{twositexcite1}
b^\dagger_{\uparrow 1} c^\dagger_{\downarrow 2} | 0 \rangle , \ \ \
b^\dagger_{\uparrow 1} b^\dagger_{\downarrow 2} | 0 \rangle \; ; 
\end{equation}
the other six  are obtained by exchanging $b$ with $c$ or 
$\uparrow$ with $\downarrow$. Two more eigenstates with two $\downarrow$-spin 
excitations are  
\begin{equation}
c^\dagger_{\downarrow 1} b^\dagger_{\downarrow 2} | 0 \rangle, \ \ \
b^\dagger_{\downarrow 1} c^\dagger_{\downarrow 2} | 0 \rangle.
\label{twositeexcite2}
\end{equation}
These ten states, as well as two other states which are not as simple to 
write down, are separated by $O(J) + O(U)$ gaps 
from the remaining four zero-energy ground states. 
The existence
of a gap for spin excitations means that the transport Green's function,
Eq.~\ref{tGfun}, 
decays exponentially in the $\tau$ direction.  Physically this 
may be interpreted\cite{Leon} as the 
expected one-dimensional Anderson localization which is driven by 
backscattering among the two counter-propagating edge states. 

Surprisingly, the problem of more than two sites can be solved exactly
in the $U \gg J$ limit.
To see this we consider the $\uparrow$-Hilbert subspace with just 
two states per site which may be represented as 
\begin{eqnarray}
b^\dagger_\uparrow | 0 \rangle &\rightarrow&  | 0 \rangle    
\nonumber \\
c^\dagger_\uparrow | 0 \rangle &\rightarrow& f^\dagger | 0 \rangle \; ,  
\label{spinless}
\end{eqnarray}
where the $f$'s correspond to spinless fermions. 
States with more than one fermion per site are forbidden by the Pauli exclusion
principle; those with more than one boson per site are disfavored 
energetically by $H_\eta$ and the first term in $H_J$, Eq.~\ref{newJ}.  
The elimination of the (zero in energy)
vacuum state is justified {\em a posteriori} as the 
ground state energy will turn out to be negative; see Eq.~\ref{dispersion}
below. The different operators appearing in the Hamiltonian (Eq.~\ref{newJ}) 
projected onto the subspace spanned by $| 0 \rangle$ and 
$f^\dagger | 0 \rangle$ can 
be represented solely in terms of the new fermion operators,
\begin{equation}
\label{newops}
 \begin{array}{l} b^\dagger_\uparrow b_\uparrow \rightarrow 1 - f^\dagger f \\ 
   \\ c^\dagger_\uparrow c_\uparrow  \rightarrow   f^\dagger f \nonumber
 \end{array} \ \ \ \
 \begin{array}{l} b^\dagger_\uparrow c_\uparrow  \rightarrow  f  \nonumber \\
   \\ c^\dagger_\uparrow b_\uparrow \rightarrow  f^\dagger
 \end{array} 
\end{equation}
This transformation is the analog of the Jordan-Wigner transformation 
\cite{negele} which maps the x-y spin chain into a system  of
non-interacting tight-binding  fermions. 
In this representation the Hamiltonian is quadratic, and setting $J=1$
we obtain:       
\begin{equation}
H_J = - \sum_{n=1}^L (1 + (-1)^n R) \: 
\big{[} f^\dagger_n f_n + f^\dagger_{n+1} f_{n+1} 
+ (-1)^n (f^\dagger_n f^\dagger_{n+1} + f_n f_{n+1}) \big{]}\ . 
\label{Uinf}
\end{equation}
The Hamiltonian is now easily diagonalized.  Due to the alternating term in 
Eq.~\ref{Uinf} it is convenient to introduce separate fermion operators 
for even and odd sites, 
\begin{eqnarray}
f_{2n} &=& d_n 
\nonumber \\
f^\dagger_{2n-1} &=& e_n \ , 
\label{unitcell}
\end{eqnarray}
where on the odd sites we have performed yet another particle-hole 
transformation. 
With this substitution, and upon Fourier transforming to $k$-space, 
the Hamiltonian may be rewritten as:
\begin{equation}
H_J = \sum_{k} \big{[} 2 d^\dagger_k d_k  - 2 e^\dagger_k e_k
+ (1 - R) (d^\dagger_k e_k - e^\dagger_k d_k)
+ (1 + R) (e^{i k}~ d^\dagger_k e_k - e^{-i k}~ e^\dagger_k d_k) \big{]}\ .
\label{kspaceH}
\end{equation}
Diagonalization is now a simple matter of computing the eigenvalues of a
$2 \times 2$ non-Hermitian matrix, and we obtain:
\begin{equation}
\epsilon_k = \pm 2 \sqrt{1 - R^2}~ |\sin(k/2)|\ .
\label{dispersion}
\end{equation}
The eigenvectors are easily found.  The $k = 0$ piece of the Hamiltonian is
defective, and once again the defect is lifted with the introduction of 
small negative $\mu_c$. 

The ground state is constructed by filling up all of the negative energy
states with the spinless fermions in the reduced Brillouin zone
$0 \leq k < \pi$.
Two things are remarkable about the dispersion relation: First, no gap opens
up for $R \neq 0$.  The expected spin gap at non-zero $R$
cannot be recovered in the $U \gg J$ limit because none of the  
excitations in the reduced Hilbert space carry spin.  
Second, the dispersion at the left and right Fermi
points is linear, unlike the quadratic dispersion found for the ferromagnetic
supersymmetric chain which describes the chiral
metal\cite{Balents}.  The linear dispersion is important to recover here
because in the original plateau transition 
problem, unlike the chiral metal, the two spatial directions are equivalent.  

In Hermitian supersymmetric theories the ground state is unique and has zero
energy.  Excited states pair into sectors of even and odd numbers of 
fermions and consequently 
the partition function $Z \equiv {\rm STr}~ e^{-\beta H_{\rm SUSY}} = 1$ 
at all temperatures $1/\beta$.  Here the supertrace operator $\rm STr$ 
is simply the ordinary
trace over states with the insertion of an additional operator $(-1)^{N_c}$, 
where $N_c$ is the total fermion number operator introduced above. 
The operator $(-1)^{N_c}$
compensates for the anti-periodic boundary condition imposed on Grassman
fields in the $\tau$-direction of the corresponding 
path-integral\cite{Balents}.
 
As the ground state energy is negative for the non-Hermitian Hamiltonian
considered here, it is important to clarify how supersymmetry is manifest. 
In particular, at non-zero temperature (corresponding in the original network 
model to finite spatial extent and periodic boundary conditions in the 
y-direction) it is important to check whether the partition function
really equals unity, $Z = 1$.  We first note that each term 
in the Hamiltonian $H_{\rm SUSY} = H_U + H_J + H_\eta$ 
either does not change the 
total fermion number, or changes it by an even integer.  Therefore the many-body
Hilbert space splits into two separate sectors: one with an odd number of
fermions and $(-1)^{N_c} = -1$, the other with an even number and
$(-1)^{N_c} = 1$.  The map between these two sectors is realized by 
fermion-valued supersymmetric charges, 
\begin{eqnarray}
Q_1 &\equiv& \sum_n \bigg{[} b^\dagger_{\uparrow n} c_{\uparrow n} 
- (-1)^n c^\dagger_{\uparrow n} b_{\uparrow n} \bigg{]}\ .
\nonumber \\
Q_2 &\equiv& \sum_n \bigg{[} (-1)^n b^\dagger_{\uparrow n} c_{\uparrow n} 
+ c^\dagger_{\uparrow n} b_{\uparrow n} \bigg{]}\ .
\label{charge}
\end{eqnarray}
which are the analogues
of those found for the supersymmetric ferromagnet\cite{Balents}.

Straightforward algebra confirms that 
$[H_{\rm SUSY}, Q_1] = [H_{\rm SUSY}, Q_2] = 0$
in the $\mu_c \rightarrow 0$ limit, and also that $\{Q_1, Q_2\} = 0$.  
This holds true for $H_{\rm SUSY}$ in the finite-$U$ (Eq.~\ref{Hterms}) and 
in the infinite-$U$ (Eq.~\ref{newJ}) limits.  Therefore 
the charges $Q_1$ and $Q_2$ map one eigenstate with an odd number of 
fermions into a second state, degenerate with the first,
but containing an even number of 
fermions.  For example, two of the four degenerate eigen-states 
of the two-site problem, 
Eq.~\ref{twositestates}, have a
single fermion while the other two states, Eq.~\ref{defectivestates}, 
have a mixture of zero and two fermions.  In this
case we find: $Q_1 | \Psi^R(1) \rangle = Q_1 | \Psi^R(2) \rangle \propto 
(|\Psi^R(3+) \rangle - |\Psi^R(3-) \rangle)$.  Operating with the charges 
a second time, on states $|\Psi^R(3\pm) \rangle$, then returns states
$|\Psi^R(1) \rangle$ and $|\Psi^R(2) \rangle$.  In this way states which differ
in content by one fermion are paired and also grouped into degenerate
quartets.  The only exception is the unique, zero-energy, empty vacuum state,
$| 0 \rangle$, which is annihilated by both supersymmetric charges, 
$Q_1 | 0 \rangle = Q_2 | 0 \rangle = 0$.  
All the other states (with either positive or 
negative\footnote{Note that $H \neq Q^\dagger Q$ as it would for a
{\bf Hermitian} supersymmetric
field theory; hence, negative energy states are permitted.} energy) 
pair off and $Z = 1$ at any temperature 
as expected from  the supersymmetry of  the construction.  
For the remainder of this paper, however, we consider only zero-temperature 
properties.

\section{DMRG Calculation}
\label{sec:dmrg}
To make further progress in the physically relevant finite-$U$  case 
we employ the Density Matrix
Renormalization Group (DMRG) introduced by White\cite{White}.  
We utilize the simpler and faster ``infinite size'' algorithm.  
It permits accurate
evaluation of observables in the thermodynamic limit of large chain length
but is less accurate for finite size chains.    

As the Hamiltonian is non-Hermitian, both right and left eigenstates 
must be found.  We employ the Davidson-Liu 
algorithm\cite{Davidson} extended by Morgan\cite{Morgan} to non-symmetric
matrices.  Sixteen on-site operators $S^a_n$, with $a = 1, 2,\cdots, 16$, 
are introduced.  In normal-ordered form these are:  
\begin{equation}
\begin{array}{l} S^1 \equiv b^\dagger_\uparrow b_\uparrow \\ \\
S^4 \equiv \bar{b}_\downarrow b_\uparrow \\ \\
S^{10} \equiv c^\dagger_\uparrow b_\uparrow \\ \\
S^{14} \equiv c^\dagger_\downarrow b_\uparrow 
\end{array} \ \ \
\begin{array}{l} S^3 \equiv b^\dagger_\uparrow b_\downarrow \\ \\
S^2 \equiv \bar{b}_\downarrow b_\downarrow + 1\\ \\
S^{16} \equiv c^\dagger_\uparrow b_\downarrow \\ \\
S^{12} \equiv c^\dagger_\downarrow b_\downarrow 
\end{array} \ \ \
\begin{array}{l} S^9 \equiv b^\dagger_\uparrow c_\uparrow \\ \\
S^{15} \equiv \bar{b}_\downarrow c_\uparrow \\ \\
S^{5} \equiv c^\dagger_\uparrow c_\uparrow \\ \\
S^{8} \equiv c^\dagger_\downarrow c_\uparrow 
\end{array} \ \ \
\begin{array}{l} S^{13} \equiv b^\dagger_\uparrow c_\downarrow \\ \\
S^{11} \equiv \bar{b}_\downarrow c_\downarrow \\ \\
S^{7} \equiv c^\dagger_\uparrow c_\downarrow \\ \\
S^{6} \equiv c^\dagger_\downarrow c_\downarrow - 1\ . 
\end{array} 
\end{equation} 
After the canonical transformation (Eq.~\ref{cantra}) the exchange part
of the supersymmetric
Hamiltonian, Eq.~\ref{Hterms}, can be represented in terms of these on-site
operators as:
\begin{equation}
H_J = -\sum_{n=1}^L~ J_n~ \bigg{[} \sum_{a=1}^8~ \lambda_a S^a_n S^a_{n+1} 
+ (-1)^n~ \sum_{a=9}^{16}~ \lambda_a S^a_n S^a_{n+1} 
\bigg{]}~ 
\label{Hamrep}
\end{equation}
with coefficients $\lambda_a = -1$ for $a = 1, 2, 10, 11, 14, 15$ 
and otherwise $\lambda_a = +1$.  
The second term in the above equation is non-symmetric; it is composed of the
eight operators ($S^a$, $a = 9, \cdots, 16$) which have fermionic character.  
When represented in terms of matrices with real-valued entries, 
the fermionic nature of these eight on-site
operators leads to additional $(-)$ signs, which   
have been taken into account in writing Eq.~\ref{Hamrep}.
In particular, these $(-)$ signs are dictated by the fermion  
ordering convention  
along the chain which we adopt to define the many-body basis states: 
$c^\dagger_i \cdots c^\dagger_j \cdots c^\dagger_k \cdots
| 0 \rangle$  with $i < j < k$. 
As the Hamiltonian commutes with the number operator on each site, 
${\cal N}_n \equiv {\cal N}^c_{\uparrow n}
+ {\cal N}^c_{\downarrow n} + {\cal N}^b_{\uparrow n} + 
{\cal N}^b_{\downarrow n}$,
the on-site operators $S^a_n$ may be represented as $4 \times 4$ 
matrices acting on the 4-dimensional single-site Hilbert space 
\begin{equation}
\begin{array}{l} | 1 \rangle \equiv b^\dagger_\uparrow | 0 \rangle \\ \\
| 2 \rangle \equiv \bar{b}_\downarrow | 0 \rangle
\end{array} \ \ \ \ 
\begin{array}{l} | 3 \rangle \equiv c^\dagger_\uparrow | 0 \rangle \\ \\
| 4 \rangle \equiv c^\dagger_\downarrow | 0 \rangle\ .
\end{array}
\label{Hilbert4}
\end{equation}
For example, the only non-zero matrix element of $S^1$ is 
$\langle 1 | S^1 | 1 \rangle = 1$.  Note that the state $\langle 2 | 
= \langle 0 | (\bar{b}_\downarrow)^\dagger = - \langle 0 | b_\downarrow$ but
the $(-)$ sign may be dropped as it always appears in pairs in the exchange 
Hamiltonian Eq.~\ref{Hamrep}.  We note that down spins are energetically 
disfavored due to the $\pm 1$ terms appearing in the definition of 
operators $S^2$ and $S^6$.  The constant terms originate 
in the canonical transformation, Eq.~\ref{cantra}. 

As in the analytically tractable $U \gg J$ limit, the ground state 
remains four-fold degenerate in the $\mu_c \rightarrow 0$ limit even down to
$U = 0$. This is accounted for by the existence of two supersymmetric 
charges, Eq.~\ref{charge}, that commute with the Hamiltonian.
The ground state expectation values of observables $\cal O$, which 
correspond to thermal expectations in the $\beta\to\infty$ limit, are 
computed using the supertrace formula,
\begin{equation}
\langle {\cal O} \rangle \equiv
\sum_{a = 1}^4 \langle \Psi^L(a) | (-1)^{N_c}~ {\cal O} 
| \Psi^R(a) \rangle~ .
\label{expect}
\end{equation}
That both left and right eigenstates appear in Eq.~\ref{expect} has its origin
in the ``resolution of identity'' which is inserted into the partition 
function for the purpose of computing expectation values:
\begin{equation}
1 = \sum_a | \Psi^R(a) \rangle~ \langle \Psi^L(a) |\ ;
\end{equation}
here the sum runs over the entire Hilbert space. (The states 
span the Hilbert space because the defect was eliminated via the introduction 
of $\mu_c < 0$.)   

We introduce reduced density matrices for both of the augmented blocks, 
each of Hilbert space size $4 M$, by 
computing partial traces over states on the right (or left) halves of the
chain.  The reduced density matrix for the left half of the chain
is given by:
\begin{equation}
\rho_{i j} = {{1}\over{8}}~ \sum_{a=1}^4~ \sum_{i^\prime=1}^{4 M}~ 
\bigg{[} {{\Psi^R_{i i^\prime}(a)~ \Psi^R_{j i^\prime}(a)}\over{
\langle \Psi^R(a) | \Psi^R(a) \rangle}} +
{{\Psi^L_{i i^\prime}(a)~ \Psi^L_{j i^\prime}(a)}\over{
\langle \Psi^L(a) | \Psi^L(a) \rangle}} \bigg{]}~ ; 
\label{density}
\end{equation}
a similar formula defines the density matrix for the right half of the chain.
Here $\Psi^{L,R}_{i i^\prime} \equiv \langle i, i^\prime | \Psi^{L,R} \rangle$
are the matrix elements of the many-body wavefunction projected onto a basis
of states labeled by unprimed roman index $i$ which covers the left half 
of the chain and primed index $i^\prime$ which covers the right half.
The states are normalized in Eq.~\ref{density} 
to insure that ${\rm Tr} \rho = \sum_i \rho_{i i} = 1$.  As can be verified
numerically, all of the eigenvalues of $\rho$ are real (as it is a
symmetric matrix) and, more significantly, they are positive 
and hence may be interpreted as probabilities.
The DMRG truncation procedure thus corresponds to the elimination 
of the least-important sectors of the Hilbert space.  Our choice for the
reduced density matrix agrees with that of Bursill, Xiang, and 
Gehring\cite{DMRG2} but differs from that of Nishino\cite{DMRG1} who argues
that a product of left with right eigenstates is optimal in a variational
sense.  However, the resulting reduced density matrix is non-Hermitian and
some of its eigenvalues are negative; they
cannot be interpreted as probabilities. Further, it is a difficult numerical 
problem to find a large fraction of the eigenstates of a non-Hermitian
matrix.  Since the reduced density matrix is introduced solely for the purpose
of truncating, in a systematic manner, the rapid growth of the Hilbert space 
with chain length, here we adopt the symmetric choice 
(Eq.~\ref{density}) and check that observables do not change
significantly when the size $M$ of the block's Hilbert space is enlarged. 

The exactly solvable $U \gg J$ limit permits another important check 
on the quality of the DMRG algorithm.  We have verified that the DMRG 
procedure accurately reproduces the energies of the 
four degenerate
ground states obtained from the dispersion relation (Eq.~\ref{dispersion})
as the chain grows.  For example, in the case $U = 10 J$ and for a 
block Hilbert space size of $M = 80$, the energies agree to within
1 part in $10^4$ at chain lengths up through $L = 16$.

We continue to lift the degeneracy, and the defectiveness, of the Hamiltonian
via a small chemical potential for fermions, Eq.~\ref{chempot}. 
Typically we set $\mu_c = -1.0 \times 10^{-6}$ in units where $J = 1$.  
Although each of the four terms in Eq.~\ref{expect} in general
depends strongly on
$\mu_c$ we have verified that this dependence drops out in the sum over the
four nearly degenerate ground states (see Fig.~\ref{bondfluct}).    
The rest of the DMRG algorithm is standard:  During each iteration, 
the chain is cut into left and
right pieces, the reduced density matrix for each half
is formed and diagonalized, new representations of the operators are found, 
and finally two sites are added to the middle of the chain and the resulting
new Hamiltonian is diagonalized via the modified Davidson-Liu algorithm.
The near degeneracy and defectiveness of $H_J$ slows down the diagonalization 
considerably 
and can be partly compensated by increasing the size of the Davidson subspace. 
We run the double-precision, vectorized, multi-processor C-code on a 
Cray EL-98 computer.  The largest computations take about 1 Gbyte of memory
and require several weeks to run. 

\section{DMRG Results}
\label{sec:results}

Previous work on the replica limit of the Heisenberg\cite{Dung-Hai} and 
Hubbard\cite{Ziqiang} models focused on the scaling of the spin gap 
$\Delta_s(R)$ as a function
of $R$, the staggered coupling between neighboring sites.  It is difficult
to gain direct access to the spin gap in the supersymmetric chain due to 
the plethora of low-lying excitations in the pure spin-up sector (discussed in
Sec. \ref{sec:infiniteU}) which persist at finite-$U$.  To isolate
spin-flip excitations would require finding many low-lying eigenvectors, 
a problem for which the DMRG method is not well suited\cite{DMRG2}.   

Critical behavior is evident, however, in the expectation value of a
spin-flip operator which lives on the bonds between neighboring sites. 
It is important to note though that according to Eq.~\ref{expect}  
supersymmetry-invariant operators (those  which commute with the 
supersymmetric charges $Q_{1,2}$),  have zero expectation value.  For example, 
$\langle H_{\rm SUSY} \rangle = 0$ despite the fact that the ground state 
energy is negative.  The non-supersymmetric fermion spin-flip operator 
$(-1)^{N_c}
S^7_n S^7_{n+1}$, on the other hand, is an appropriate choice.  
Even with no applied dimerization ($R = 0$), 
its ground-state expectation value 
\begin{equation}
\langle (-1)^{N_c}~ S^7_n~ S^7_{n+1} \rangle 
= \langle (-1)^{N_c}~ c^\dagger_{\uparrow n} c_{\downarrow n} 
c^\dagger_{\uparrow n+1} c_{\downarrow n+1} \rangle 
\equiv B(n, n+1)
\label{bondop}
\end{equation}
has even/odd modulation due to the free boundary condition at the chain ends.  
This modulation can be isolated by defining
\begin{equation}
\Delta(n) \equiv (-1)^{n} [B(n, n+1) - B(n-1, n)] + c.c. \ ,  
\label{Delta}
\end{equation}
and it can be understood as follows: chains with an even number of sites
and free boundary conditions at the chain ends
have no reflection symmetry about any site center.  The chain ends
induce slight dimerization (which also breaks reflection symmetry about site 
centers).  Each time the chain length is increased by two sites during a DMRG
iteration the pattern of weak and strong bonds alternates; the term
$(-1)^{n}$ appearing in Eq.~\ref{Delta} accounts for this alternation.    

As $\Delta(n)$ is the expectation value of an operator which 
induces dimerization at $R \neq 0$ 
it has scaling dimension $x_\Delta$ which is related to the 
correlation length exponent by the standard scaling relation: 
$\nu = (2 - x_{\Delta})^{-1}$\cite{cardybook}.  Therefore, in a finite 
system we expect
\begin{equation}
\label{scaling}
\Delta(n) \sim L^{-x_{\Delta}}, \ \   L \rightarrow \infty .
\end{equation}
This observation provides us with a means
of extracting $x_{\Delta}$ from the DMRG by measuring  the 
expectation 
value of the dimerization operator on the central bonds, $\Delta(L/2)$,
at each DMRG step.  

\begin{figure}
\epsfxsize = 10cm \epsffile{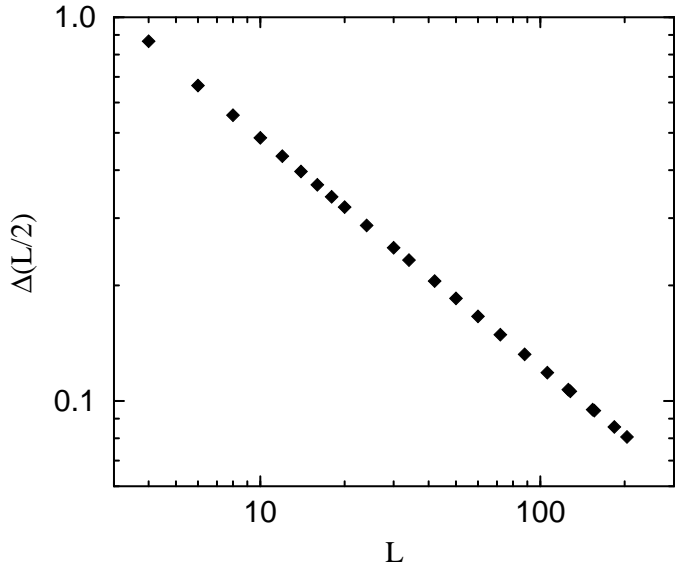}
\caption{
\label{bondfluct2} 
Expectation value of the dimerization operator on the central 
bonds, $\Delta(L/2)$, for the $S=1/2$ Heisenberg spin chain; $L$ is 
the chain length.  The infinite-size DMRG method with a block 
Hilbert space size of $M = 256$ is used. Slight curvature of the plot signals 
the presence of other less relevant scaling operators and logarithmic 
corrections to scaling.
}
\end{figure}

To test this method, and the DMRG algorithm,
we first examine the ordinary isotropic, nearest-neighbor, spin-$1/2$ SU(2) 
quantum antiferromagnet.  In this case the dimension of the 
dimerization operator $(-1)^n~ \vec{S}_n \cdot \vec{S}_{n+1}$
is known exactly\cite{aff}: $x_{\Delta} = 1/2$.    
For chains up to length $L = 204$ and block Hilbert space
size $M = 256$ we find results consistent with the theoretical 
prediction.  Namely, if we calculate $x_\Delta$ by 
fitting the log-log plot in Fig.~\ref{bondfluct2} to a linear function, for 
points $L > 10$, we find $x_\Delta \approx 0.6$. The discrepancy  
with the exact result we attribute to the presence of a marginally irrelevant
operator\cite{aff},  
and other less relevant scaling operators. These corrections to 
scaling, which lead to a slightly larger value of the effective exponent, are 
evidenced by a slight curvature of the plot in  Fig.~\ref{bondfluct2}. 
More precisely, the marginally irrelevant operator makes multiplicative 
logarithmic corrections to scaling\cite{cardybook}. We take these corrections
into account by fitting the values to the functional form 
$\Delta(L/2) = L^{-x_\Delta}(\log(L))^\gamma f(L^{-1})$.  
The function $f(x)$ is assumed analytic around the origin and we
expand it via Taylor series, keeping only the first two terms,  
$f(x) = C_1 + C_2x $. With this procedure we obtain
$x_\Delta \approx 0.5$ and $\gamma \approx 0.5$, in agreement 
with theoretical expectations. 

\begin{figure}
\epsfxsize = 10cm \epsfbox{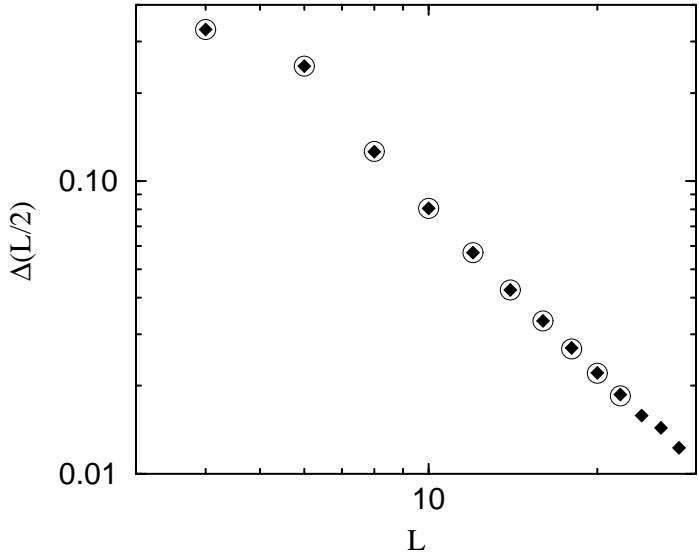}
\caption{
\label{bondfluct}
Expectation value of the dimerization operator on the central bonds,
$\Delta(L/2)$, versus system length  
$L$ for $U = 0$ and at the critical point $R = 0$ of the SUSY chain.  
Results for two different runs are shown: filled diamonds are for
a block Hilbert space size of $M = 160$ [corresponding to a total Hilbert
space size of $(4M)^2 = 409,600$] and $\mu_c = -10^{-6}$ in units where
$J = 1$.  Open circles are for a more accurate calculation 
at $M = 256$ [corresponding to a total Hilbert space size of $1,048,576$]
and $\mu_c = -10^{-4}$.
The two calculations agree precisely up through chain length $L = 8$. 
For $L > 8$ the entire Hilbert space cannot be accommodated; instead it is
truncated via the DMRG algorithm.  Nevertheless, the two calculations continue
to agree closely out through chain length $L = 22$, the largest chain length
reached in the $M = 256$ calculation.  At $L = 26$ the $M = 160$ point
deviates from the expected scaling form.  We attribute the breakdown of 
scaling at large chain length to the truncation of the Hilbert space. 
} 
\end{figure}

Returning to the supersymmetric chain,
scaling is evident in Fig.~\ref{bondfluct} where 
$x_\Delta \approx 1.8$ is calculated by fitting the $M = 160$ 
points with $L > 10$ to a straight line; 
curvature in the plot signals once again corrections to scaling. 
We account for these less relevant operators
by fitting $\Delta(L/2)$ to the functional form 
$\Delta(L/2) = L^{-x_\Delta} f(L^{-1})$.  As in the case of the $SU(2)$
Heisenberg chain above, 
we Taylor-expand $f(x) = C_1 + C_2 x + \cdots$, keeping only the 
first two terms. Following this procedure we obtain $x_\Delta = 1.58 \pm 0.07$.
The error is estimated by choosing different sets of points to 
include in the 
fit; statistical errors are smaller, at the few percent level.  
This value of the scaling dimension is consistent with the more precise value 
$x_\Delta = 2 - \nu^{-1} = 1.57 \pm 0.01$ (using $\nu = 2.33 \pm 0.05$), 
reported by other groups\cite{numerics}.  

The DMRG works best for phases with a gap to low-lying excitations; we do
not expect, nor do we find, very good quantitative agreement at the 
critical point. 
To obtain more accurate scaling it would be necessary to employ the 
``finite-size'' DMRG algorithm and carry out careful finite-size analysis
by varying both the chain length $L$ and the dimerization 
parameter $R$ to search for a universal scaling function.
 
\section{Discussion}
\label{sec:discussion}
What have we learned from the above analysis?  First,
the scaling in Fig.~\ref{bondfluct} demonstrates the 
utility of the supersymmetric approach to understanding transport at the 
IQHE plateau transition.
The series of transformations outlined above which map the network model 
described by Eq.~\ref{hamilton} to the pure, but rather complicated, 
supersymmetric spin-chain Hamiltonian Eq.~\ref{Hamrep} appears to work. 
The replica trick is avoided and we obtain instead a concrete model amenable
to analytical and numerical study.  Deep inside a plateau ($R \rightarrow 
\pm 1$) the chain breaks up into disconnected two-site systems and 
Anderson localization in the interior is complete.  At $R = -1$
the two spins at the chain ends are fully polarized in the 
$\uparrow$-direction and no extended
edge currents flow; in the opposite $R = +1$ limit the two spins are free
and the conductivity increases by a quantum\cite{Kim}. 

Second, and more importantly, 
the Hamiltonian is analytically tractable when projected onto the Hilbert 
space with only one type ($\uparrow$) of boson and fermion.
This is the $U \gg J$ limit, Eq.~\ref{newU},  
and there is the intriguing possibility that the Hamiltonian 
can be studied systematically down to the physical regime of finite $U$.  
This approach
differs fundamentally from previous weak-coupling studies\cite{Andreas} where
runaway RG flows toward strong coupling in the generic IQHE case suggested the
existence of a strong-coupling fixed point.  The starting point here, by
contrast, would be at very strong coupling.  It may then be 
possible to incorporate spin excitations perturbatively about the 
strong coupling limit.  

A number of unanswered questions remain.  The significance
of the defectiveness of the Hamiltonian is unclear.  It is a robust feature,
as it occurs at both infinite and finite $U$.  Also unanswered is the question
of whether or not it will be possible to construct a two-dimensional 
conformal field theory (CFT) of the critical point. As pointed out by 
Mudry {\em et al.} \cite{mudry} 
a candidate CFT would have to be non-unitary, thus 
allowing for scaling dimensions not predicted by the Kac determinant. This 
opens up the possibility of identifying a CFT that contains an operator with 
a scaling  dimension $x_\Delta\approx 11/7$ in its spectrum; such an 
operator, if associated with a perturbation that induces a flow from the 
critical point to the non-critical phases of two adjacent Hall plateaus, 
would give rise to the expected  correlation length exponent 
$\nu=(2-x_\Delta)^{-1}\approx 7/3$. Searching for such a field theory among
CFT's which are also supersymmetric has not been successful~\cite{wen}.  
We believe that further studies of the supersymmetry 
approach to  the Chalker-Coddington network model will  shed 
more light on this matter.

\section{Acknowledgments}
We thank L. Balents, R. Bursill, D. Nelson, N. Read, D. Serban,  
R. Shankar, and X.-G. Wen 
for helpful discussions, and W. Becker for pointing out
references on the diagonalization of large matrices.  
Some of our work was done at the Aspen Center for
Physics.  Computations were carried out at the Theoretical Physics
Computing Facility at Brown University.  This work was supported in part 
by the National Science
Foundation through Grants Nos. DMR-9313856 and No. DMR-9357613 and by 
a grant from the the Alfred P. Sloan Foundation (J.B.M.).

\end{document}